\documentstyle[multicol,epsf,aps]{revtex}
\begin{document}
\draft
\title{Search for Solar Axions Using $^{57}$Fe}
\author{M. Kr\v{c}mar, Z. Kre\v{c}ak, M. Stip\v{c}evi\'{c}, and
       A. Ljubi\v{c}i\'{c}}
\address{Rudjer Bo\v{s}kovi\'{c} Institute, POB 1016, 10001 Zagreb,
            Croatia}
\author{D.A. Bradley}
\address{Department of Physics, University of Malaya, 50603 Kuala
       Lumpur, Malaysia}
\date{\today}
\maketitle

\begin{abstract}
    We have made a search for $^{57}$Fe gamma rays of  energy  14.4~keV
    induced by resonant absorption of monochromatic solar axions,
    as proposed by Moriyama. The proposed axions are suggested to be
    emitted from the Sun, in M1 transitions between the first, thermally
    excited state and the ground state of $^{57}$Fe. An upper limit on
    hadronic axion mass of  745~eV is obtained at the 95\%
    confidence level, it being assumed that $z=0.56$ and $S=0.5$.
\end{abstract}

\begin{multicols}{2}

The existence of axions, proposed as very light neutral pseudoscalar
bosons which couple weakly to stable matter, would solve the strong
CP problem
associated with the $\Theta$--vacuum structure of QCD. The effects
of  $\Theta$--vacuum lead to an effective interaction which
violates CP invariance in strong interactions, the magnitude of the
interaction being proportional
to the parameter $\Theta$. A limit on $\Theta$ of less than (or of
an order of) 10$^{-9}$ follows from present experimental bounds
upon the electric dipole moment of the neutron. In an attempt to
guarantee strong CP invariance automatically, Peccei
and Quinn have introduced an additional global chiral symmetry,
spontaneously broken at an energy scale $f_{a}$, yielding
$\Theta$=0 dynamically \cite{Pec77}. This solution predicts the
existence of a pseudo Nambu--Goldstone boson, called the axion,
whose mass ($m_{a}$) is inversely proportional to $f_{a}$
\cite{Wei78}. Axions also arise in supersymmetric and superstring
theories and are candidates for  dark matter in the universe.
In the standard axion model \cite{Pec77,Wei78}
the Peccei--Quinn (PQ) symmetry
breaking scale is assumed to be equal to the scale of electroweak
symmetry breaking. The axion mass would be roughly  of the order
of 100~keV to 1~MeV and
has been experimentally excluded \cite{Don78}. Variant
axion models  keep $f_{a}\sim250$~GeV,
but drop the constraints of tree--level flavor conservation and
predict an axion mass near to 1.7~MeV \cite{Pec86}. The
existence of variant axions has
similarly been ruled out by experiment \cite{Bar87}. In efforts to retain 
the PQ idea, new axion models have been proposed which decouple the scale
of PQ symmetry breaking from the electroweak scale and
introduce $f_{a}$ at a value much greater
than 250~GeV. Because coupling constants of axions with matter and
radiation are inversely proportional to $f_{a}$,  axion models of this type
are generically referred to as invisible axions. The
two classes of invisible axion models which have been proposed are
referred to as KSVZ or hadronic axions
\cite{Kim79} and DFSZ or ``GUT'' axions
\cite{Din81}. While astrophysical and cosmological
considerations constrain the mass of these invisible axions to
a rather narrow range of
$10^{-5}~{\rm eV}\leq m_{a}\leq 10^{-2}$~eV, 
but with large uncertainties
on either side, less model--dependent laboratory experiments have excluded
masses greater than about 10~keV \cite{Tur90,Bar96,Raf97}.               
However, it should be noted that for hadronic axions there exists a small 
window $10~{\rm eV}\leq m_{a}\leq 20~{\rm eV}$, 
between the supernova (SN)
1987A cooling and axion burst arguments \cite{Bar96,Raf97,Eng90},  
as long as the axion--photon coupling is sufficiently small 
($E/N\approx 2$).

Because of axion coupling to the nucleus, Moriyama \cite{Mor95}
has proposed the existence of almost monochromatic axions produced
in the solar interior during M1 transitions  between the first,
thermally excited state of 14.4~keV and the ground state of
$^{57}$Fe . The stable isotope of iron, $^{57}$Fe                      
(with natural abundance 2.2\%), is exceptionally abundant among the heavy
elements in the Sun (solar abundance by mass fraction
2.7$\times10^{-5}$). The
axions are Doppler broadened due to thermal motion of the axion
emitter in the Sun and therefore they are able to excite
the same nuclide. In the laboratory it is possible that this effect could 
be measured by absorption of solar axions in a $^{57}$Fe target.
The possibility of decay of the axion into two photons during their traversal  
from the Sun to the Earth is insignificant. The resonant excitation of
$^{57}$Fe by axions would be accompanied by subsequent
decays of excited nuclei,
either through  emission of a gamma ray with an energy of
14.4~keV, or through emission of an internal conversion electron.
By detecting these decay products one could make conclusions about
the mass of the axion. This makes possible an experimental
test of the hadronic axion window, independent of the axion--photon
coupling. Experiments based on the axion--photon coupling have been proposed
and results were reported by several authors \cite{Sik83}.

Following the calculations in Ref. \cite{Mor95}, one can express
the total rate of  excitation per unit mass of $^{57}$Fe in the
laboratory as
\begin{equation}
     R_{\rm exc} = 1.65\times10^{24} 
     \left( \frac{\Gamma_{a}}{\Gamma_{\gamma}} \right) ^{2}
     {\rm day^{-1}~g^{-1}}~, 
     \label{eq1}
\end{equation}
where $\Gamma_{a}/\Gamma_{\gamma}$ represents the branching ratio
of the M1 axionic transition, relative to the gamma transition
\cite{Avi88} and contains the nuclear--structure--dependent terms
$\beta$ and $\eta$, as well as the isoscalar and
isovector axion--nucleon coupling constants $g_{0}$ and $g_{3}$.
\begin{equation}
  \frac{\Gamma_{a}}{\Gamma_{\gamma}} = \left( \frac{k_a}{k} \right)^{3}  
  \frac{1}{2\pi\alpha} \frac{1}{1 + \delta^{2}}
  \left[ \frac{g_{0}\beta + g_{3}}
  {\left( \mu_{0} - \frac{1}{2} \right)\beta + \mu_{3} - \eta} \right]^{2}~, 
  \label{eq2}
\end{equation}
where $k_{a}$ and $k$ are the momenta of the photon and the axion,
respectively,  $\alpha\simeq1/137$ is the fine structure
constant, $\delta\sim0$ is the E2/M1 mixing ratio, $\mu_{0} -
\frac{1}{2}\sim0.38$, $\mu_{3}\sim4.71$, $\beta=-1.19$ and
$\eta=0.80$. The quantities $\mu_{0}$ and $\mu_{3}$
denote the isoscalar and
isovector magnetic moments, respectively, while {\em g$_{0}$} and
{\em g$_{3}$} are related to the axion mass in the hadronic axion
model  \cite{Kap85} as follows:
\begin{equation}
 g_{0} = -7.8 \times 10^{-8}\left(\frac{m_{a}}{\rm 1~eV}\right)
       \left(\frac{3F-D+2S}{3}\right) 
       \label{eq3}
\end{equation}
and
\begin{equation}
 g_{3} = -7.8\times10^{-8}\left(\frac{m_{a}}{\rm 1~eV}\right)
         \left[\left(F+D\right)\frac{1-z}{1+z}\right]~.  
         \label{eq4}
\end{equation}
The mass of the axion is related to $f_{a}$ by
\[ m_{a} = 1~{\rm eV}\frac{\sqrt{z}}{1+z}\frac{1.3\times
                10^{7}}{f_{a}/{\rm GeV}}~. \]
The constants $F$ and $D$ are the  invariant matrix elements of
the axial current, determined from semileptonic decays, $S$
is the flavor--singlet axial--vector matrix element, and
$z=m_{u}/m_{d}\sim0.56$ is obtained in a first order
calculation of the quark mass ratio. The naive quark model gives
$S=3F-D=0.58$. However, the estimation of $z$ and
$S$ suffer from large uncertainties and ambiguity, and are still
poorly constrained parameters. The error on $z$ is of the magnitude of
a second--order correction \cite{Bar96}, while the experimental
value of $S$,  extracted from the polarized structure
function data, ranges
from $-0.09$ to 0.57 \cite{May88}. Combined reanalyses
have been carried out, from
recent data taken using different targets, including NLO
perturbative QCD \cite{Alt97,Ada97}.
A value $S\simeq 0.5$ was estimated, with $F+D=1.257$
and $F/D=0.575$.

We have searched for a peak corresponding to the 14.4~keV gamma ray
of $^{57}$Fe in
a single spectrum measured by a Si(Li) detector (SLP--10180--P, ORTEC).
The energy resolution at
this photon energy is 235~eV for presently used facilities. The total
internal conversion coefficient
for nuclear deexcitation of the first excited state of $^{57}$Fe
is 8.56 \cite{Bha92}, and thus the corresponding probability for
emission of 14.4 keV gamma rays is  $\gamma=0.105$. We have
obtained an energy calibration using the radioactive sources
$^{55}$Fe, $^{57}$Co
and $^{241}$Am (IAEA) with a diameter
of about 7~mm. The target of $^{57}$Fe was a M\"{o}ssbauer
absorber, 95\% enriched (WISSEL), with a diameter of 10~mm and a
thickness of about 53~$\mu$m. A disc of natural iron with the same
dimensions was used for background measurement.
The mass of $^{57}$Fe in the enriched target
is $M=31.53\times10^{-3}$~g, and a distance between the target
and  beryllium window of the detector of 3~mm was used.
Since the attenuation length
of the 14.4~keV gamma ray is 20~$\mu$m in iron, the average escape
probability from the target is $\xi=0.35$. We have estimated a total
efficiency for detection of 14.4~keV gamma rays  using the
13.9~keV gamma ray emission from the $^{241}$Am source
(activity  2.05$\times10^{3}$~Bq)
and obtained $\varepsilon=(1.6\pm0.1)\times10^{-2}$.

\narrowtext

\begin{figure}[h]
\begin{center}
\leavevmode
\epsfxsize=0.45\textwidth
\epsffile{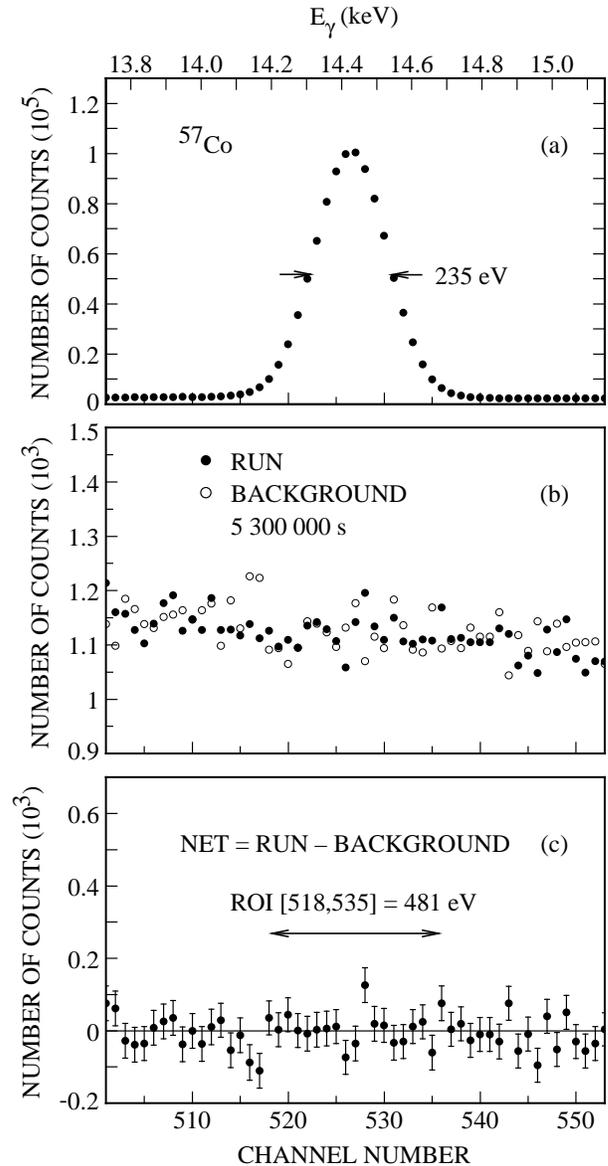}
      \caption{
      (a) The 14.4~keV gamma ray peak from the $^{57}$Co source.
      (b) Energy spectra measured with the enriched
      $^{57}$Fe (run) and with the natural iron (background)
      targets, accumulated for time periods of $5.3\times 10^{6}$
      s. (c) Net number of counts in the region of the 14.4~keV
      gamma ray peak in an effort to detect axionic excitation of
      $^{57}$Fe. \label{fig1}}
\end{center}
\end{figure}

The position and width of the peak have been estimated by measuring
the 14.4~keV gamma ray peak from the $^{57}$Co M\"{o}ssbauer source,
Fig.\ \ref{fig1}(a). Energy spectra have been measured with the
enriched $^{57}$Fe target, providing  sensitivity to monochromatic solar
axions, and with the natural iron target,
Fig.\ \ref{fig1}(b). The difference between these two spectra is shown
in Fig.\ \ref{fig1}(c). Data and background were both accumulated for
time periods $\Delta t=61.343$~days. We have detected
$N_{\gamma}=56\pm201$ gamma ray events in the energy interval of
481~eV. Equations\ (\ref{eq1})--(\ref{eq4}) allow us to relate the
axion mass to the measured value of $N_{\gamma}$,
\begin{equation}
  \left( \frac{m_{a}}{1{\rm eV}} \right)^{4} =
  \frac{4.9\times 10^{3}}{C^{4}}   
  \frac{N_{\gamma}}{M\cdot\Delta t\cdot\gamma\cdot\xi\cdot\varepsilon}~,  
  \label{eq5}
\end{equation}
where $C=\beta(3F-D+2S)/3+(F+D)(1-z)/(1+z)$. If the value of $S$
is 0.5, as suggested by recent analyses \cite{Alt97,Ada97},
we obtain $m_{a}^{4}=(4.5\pm16.0)\times10^{10}~{\rm eV}^{4}$.
To determine
an  upper limit to the invisible hadronic axion mass, we have multiplied
our statistical error by 1.645 and added in the value of $m_{a}^{4}$ to
obtain $m_{a}<745$~eV at the 95\% confidence level. It can be
noted that in
Eq.\ (\ref{eq5}) we have omitted a phase space factor $(k_{a}/k)^{3}$.
The effect of this on the measured axion mass in our experiment is
estimated to be less than $2\times 10^{-3}$. We have also considered 
self absorption of monochromatic axions with $m_{a}\leq 1$~keV in 
the solar interior, in addition to the effect of absorption by the Earth        
(day/night modulation). The mean free path of the axions has been calculated 
to be $(2\pi )^{1/2}(\sigma_{em}^{2}+\sigma_{ab}^{2})^{1/2}/\sigma n$,         
the effects of nuclear recoil ($1.9\times 10^{-6}$~keV) and the  
redshift due to the gravitation of the Sun ($1.5\times10^{-4}$~keV) being 
considered negligible. The quantities
$\sigma_{em}$ ($\sim 10^{-3}$~keV) and $\sigma_{ab}$ 
($\sim 10^{-3}$~keV$\rightarrow 10^{-5}$~keV) represent  Doppler
broadening of the 14.4~keV line of $^{57}$Fe at the temperature of the 
emitter and the absorber, respectively, while $n$ is the average density
of $^{57}$Fe atoms in the absorber.
The integrated cross section for axion resonant absorption is
$\sigma = \pi \sigma_{0}\Gamma \Gamma_{a}/\Gamma_{\gamma}$, where
$\sigma_{0}=2.6\times 10^{-18}~{\rm cm}^{2}$ is the maximum resonant      
cross section of $\gamma$ rays and $\Gamma=4.7\times 10^{-12}$~keV is 
the total decay width of the first excited state of $^{57}$Fe. From the 
above, it is indicated  that axions of mass $m_{a}\leq 
1$~keV would escape from the Sun with insignificant absorption ($<10^{-7}$).   
The day/night modulation is less than $10^{-5}$.
    
We have performed the first measurement of nuclear axion emission from
the Sun, deriving an upper limit to the hadronic axion mass of 745~eV.
The SN 1987A bound for hadronic axion mass is 20~eV \cite{Eng90}. Note however 
that both limits depend upon astrophysical and particle 
physics parameters. Solar axion emission has been estimated using temperature
and  $^{57}$Fe density distributions, obtained within the framework of the
standard solar model \cite{Tur93}. SN 1987A axion emission was
estimated, identifying  absorption probability for a given coupling
to nucleons ($g_{aN}$) within a simplified model of the supernova density 
and temperature profile. However, statistics dominate the uncertainties  
associated with the axion burst argument, with only 19 neutrinos being observed 
from SN 1987A. Observations of other supernovae will be required if the 
uncertainty is to be reduced. Recent measurements of the particle physics
parameters, used in obtaining the SN 1987A upper mass boundary, allow 
estimation of expected events in the Kamiokande II study.
{}From Fig. 1 in Ref. \cite{Eng90} we have scaled the events,
expected due to absorption of 
10.96~MeV axions into the $0^{-}$ state of $^{16}$O, using $c_{0}^{2}$, 
where $c_{0}=(3F-D+2S)/6$ is the dimensionless  isoscalar coefficient. From 
the derived value of $g_{aN}\approx 2\times 10^{-6}$ and the recent value of
$g_{aN}=5\times 10^{-8}m_{a}/(1{\rm eV})$ we have recalculated the upper limit
of hadronic axion mass from the axion burst argument to be $m_{a}\leq 40$~eV. 
   
We conclude, by noting that improvements in detection of 14.4~keV gamma
rays from $^{57}$Fe atoms are possible, by seeking increase in the value
of the factor $M\cdot \xi \cdot \varepsilon$
in Eq. (5) and by further suppressing the background. This new method of
investigation appears very promising in determining the 
hadronic axion window, being independent 
of  supernova models and the uncertainties associated with them.

The authors wish to thank the Ministry of Science and Technology of Croatia
for financial support. One of the authors (D.A.B.) is grateful for the  
support of the University of Malaya.

\end{multicols}

\end{document}